\documentclass[12pt]{article}

\usepackage{amsmath,amssymb,amsthm,enumitem}
\usepackage{hyperref}
\usepackage{geometry}
\usepackage{subcaption}
\usepackage{tikz}
\usetikzlibrary{shapes,arrows,calc,positioning}

\title{\textbf{JANUS: A Stablecoin 3.0 Blueprint for Navigating the Stablecoin Trilemma Through Dual-Token Design, Multi-Collateralization, Soft Peg, and AI-Driven Stabilization}}
\author{Stylianos Kampakis, PhD, CStat \\
\texttt{stelios@janusdefi.com, stelios@thedatascientist.com}}
\date{\today}

\theoremstyle{definition}

\begin{document}

\maketitle

\begin{abstract}
This paper introduces JANUS, a Stablecoin 3.0 protocol designed to address the stablecoin trilemma—simultaneously improving decentralization (D), capital efficiency (E), and safety/stability (S). Building upon insights from previous stablecoin generations, JANUS leverages a dual-token system (Alpha and Omega), integrates crypto-assets and real-world assets (RWAs), employs a soft-peg mechanism, and utilizes AI-driven stabilization.

We provide a comprehensive theoretical framework, including formal definitions of D, E, and S, along with equilibrium existence proofs and analogies drawn from international trade and open-economy macroeconomics. By introducing a second token backed by external yield, JANUS breaks from ponzinomic dynamics and creates a more robust foundation. Multi-collateralization and a soft peg enable controlled price oscillations, while AI-driven parameter adjustments maintain equilibrium. Through these innovations, JANUS aims to approach the center of the stablecoin trilemma, offering a globally resilient, inflation-adjusted, and decentralized stablecoin ecosystem bridging DeFi and TradFi.

The main body presents a high-level overview of the trilemma and JANUS’s key features, while the Appendix provides more formal mathematical treatments, including rigorous metrics for decentralization, capital efficiency, and stability, as well as the optimization challenges inherent in the trilemma.
\end{abstract}

\tableofcontents
\newpage

\section{Introduction and Conceptual Overview}
Stablecoins play an essential role in decentralized finance (DeFi), enabling frictionless on-chain transactions, hedging strategies, and bridging value between blockchain networks and traditional financial systems. Over the years, stablecoins have evolved through multiple design paradigms, each bringing benefits and encountering new challenges. Despite these innovations, trade-offs remain among decentralization, capital efficiency, and safety/stability—collectively known as the stablecoin trilemma.

\subsection{Motivation and Key Features of JANUS}
Early stablecoins (\textbf{Stablecoin 1.0}) were fully backed by fiat reserves, achieving short-term stability and capital efficiency but relying on centralized custodians, thereby limiting decentralization. Later designs (\textbf{Stablecoin 2.0})—such as algorithmic and inflation-indexed stablecoins—introduced mechanisms to preserve long-term purchasing power (e.g., flatcoins, reflex-indexed assets). However, they often failed to excel across all three key dimensions:
\begin{itemize}[noitemsep]
    \item Minimizing trusted intermediaries and single points of failure (\textbf{decentralization}).
    \item Avoiding heavy overcollateralization and excess capital lockup (\textbf{capital efficiency}).
    \item Maintaining robust price stability and a negligible run probability (\textbf{safety/stability}).
\end{itemize}

To address these shortcomings, we propose \textbf{JANUS}, a Stablecoin 3.0 framework integrating four primary features:
\begin{enumerate}[noitemsep]
    \item \textbf{Dual-Token Architecture:} Introducing a second token (Omega), backed by RWAs and yielding external returns, mitigating ponzinomic dynamics.
    \item \textbf{Multi-Collateralization:} Combining crypto assets and tokenized real-world assets to reduce volatility and systemic risk.
    \item \textbf{Soft-Peg Mechanism:} Permitting controlled oscillations around a reference price rather than a strict $1{:}1$ parity, thereby reducing fragility.
    \item \textbf{AI-Driven Stabilization:} An automated feedback loop that adjusts fees, rewards, and vault parameters to maintain equilibrium amid evolving market conditions.
\end{enumerate}

By combining these, JANUS targets a more balanced approach to the stablecoin trilemma—achieving higher \emph{decentralization} (D), \emph{capital efficiency} (E), and \emph{safety/stability} (S) simultaneously.

\subsection{The Stablecoin Trilemma and Its Formal Metrics}
The stablecoin trilemma posits that no single design can fully maximize D, E, and S without compromise. Historically, stablecoins have found themselves optimizing two of the three while sacrificing the third:
\begin{itemize}[noitemsep]
    \item \emph{Fiat-backed stablecoins} deliver short-term stability and capital efficiency but rely on centralized custodians, weakening D.
    \item \emph{Crypto-backed stablecoins} maintain higher D but require heavy overcollateralization, harming E, and can be vulnerable to correlated crypto market downturns, impacting S.
    \item \emph{Algorithmic stablecoins} chase decentralization and capital efficiency but, lacking external anchors, risk major price failures (low S) under adverse market sentiment.
\end{itemize}

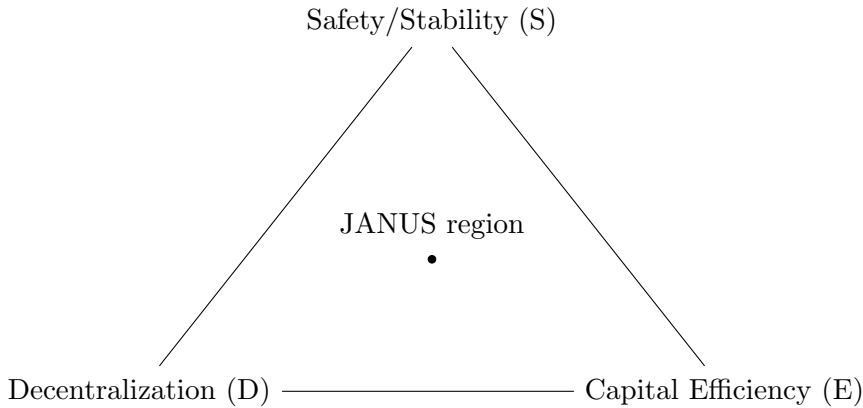
\begin{figure}[h!]
\centering
\begin{tikzpicture}[scale=3.5, every node/.style={font=\small}]
\node (D) at (0,0) {Decentralization (D)};
\node (E) at (2.2,0) {Capital Efficiency (E)};
\node (S) at (1.1,1.4) {Safety/Stability (S)};
\draw (D) -- (E) -- (S) -- (D);

\node[draw, circle, inner sep=1pt, fill=black] (P) at (1.1,0.5) {};
\node[above=2pt of P] {JANUS region};
\end{tikzpicture}
\caption{The stablecoin trilemma: It is challenging to optimize decentralization, capital efficiency, and safety/stability at the same time. JANUS strives to occupy a region closer to the triangle’s center, balancing all three dimensions.}
\label{fig:trilemma_visual}
\end{figure}

\paragraph{Defining Decentralization, Capital Efficiency, and Safety/Stability.}
In the Appendix, we present rigorous mathematical definitions for each dimension:
\begin{itemize}[noitemsep]
    \item $D(U)$: A measure of decentralization (e.g., via Herfindahl-like metrics of governance distribution). Higher $D(U)$ indicates more balanced control and minimal trust requirements.
    \item $E(U)$: A capital efficiency index, typically the ratio of stablecoin supply (at target price) to total collateral value. Lower overcollateralization implies higher $E(U)$.
    \item $S(U)$: Safety/stability modeled as $1 - \mathbb{P}(\mathcal{F})$, where $\mathbb{P}(\mathcal{F})$ is the probability of catastrophic failure, such as a prolonged peg break.
\end{itemize}

While no stablecoin design can perfectly maximize D, E, and S, JANUS’s dual-token model, multi-collateralization, soft peg, and AI-driven stabilization collectively \emph{expand the feasible frontier}, yielding higher $(D(U), E(U), S(U))$ than previous solutions. The next section details how these components reduce systemic fragility and foster robust, inflation-adjusted value, culminating in a non-ponzinomic equilibrium.

\subsection{Ponzinomic Pitfalls in Existing Protocols}

A ponzinomic stablecoin relies primarily on new inflows to sustain price appreciation, lacking a fundamental source of external value. Once fresh buyers diminish, the token collapses because there are no uncorrelated yields or real reserves to support its price. Many existing stablecoins partially exhibit such vulnerabilities. Below is a brief comparative assessment of some well-known stablecoins (both legacy and experimental) and their respective positions along decentralization, capital efficiency, safety/stability, and ponzinomic risk.

\begin{table}[h!]
\centering
\small % Even smaller than \small
\setlength{\tabcolsep}{3pt} % Reduce horizontal padding
\renewcommand{\arraystretch}{1.0} % Adjust vertical spacing
\begin{tabular}{lcccc}
\hline
\textbf{SC} & \textbf{Decent.} & \textbf{Cap. Eff.} & \textbf{Safety} & \textbf{Ponzi Risk}\\
\hline
\textbf{DAI} & Med & Low--Med & Med & Low \\
\textbf{USDT} & Low & High & Med & Very Low \\
\textbf{USDC} & Low & High & High & Very Low \\
\textbf{UST} & Med & High & Low & Very High \\
\textbf{Flatcoins} & Med--High & Med & Med--Low & Med--High \\
\hline
\end{tabular}
\caption{High-level comparison of select stablecoins. ``Decent.'' = Decentralization, ``Cap. Eff.'' = Capital Efficiency, ``Safety'' = Safety/Stability. Ponzi risk rises when external anchors are weak or absent, and price depends heavily on new inflows.}
\label{tab:stablecoin_comparison}
\end{table}

In \autoref{tab:stablecoin_comparison}, we see that:
\begin{itemize}[noitemsep]
    \item \textbf{USDT and USDC} (centralized fiat-backed) score high on capital efficiency and safety in normal conditions but compromise heavily on decentralization.
    \item \textbf{DAI} has moderate decentralization but can require overcollateralization (weakening capital efficiency). Recent reliance on centralized stablecoins as part of its collateral has further diluted D.
    \item \textbf{UST (TerraUSD)} before collapse exemplified how purely algorithmic approaches lacking external anchors can suffer catastrophic failures under market stress, revealing very high ponzinomic risk.
    \item \textbf{Flatcoins} aim for inflation hedging but typically remain reliant on continuous buy demand for price support, raising ponzinomic concerns if external yield sources or uncorrelated collateral are absent.
\end{itemize}

JANUS aims to avoid these pitfalls by combining multiple forms of collateral, introducing an external yield token (Omega), and allowing controlled price deviations that do not hinge solely on reflexive demand.

\subsection{The Stablecoin Trilemma and Its Formal Metrics}

The stablecoin trilemma posits that no single design can fully maximize D, E, and S without compromise. Historically, stablecoins have found themselves optimizing two of the three while sacrificing the third:
\begin{itemize}[noitemsep]
    \item \emph{Fiat-backed stablecoins} deliver short-term stability and capital efficiency but rely on centralized custodians, weakening D.
    \item \emph{Crypto-backed stablecoins} maintain higher D but require heavy overcollateralization, harming E, and can be vulnerable to correlated crypto market downturns, impacting S.
    \item \emph{Algorithmic stablecoins} chase decentralization and capital efficiency but, lacking external anchors, risk major price failures (low S) under adverse market sentiment.
\end{itemize}

\begin{figure}[h!]
\centering
\begin{tikzpicture}[scale=3.5, every node/.style={font=\small}]
\node (D) at (0,0) {Decentralization (D)};
\node (E) at (2.2,0) {Capital Efficiency (E)};
\node (S) at (1.1,1.4) {Safety/Stability (S)};
\draw (D) -- (E) -- (S) -- (D);

\node[draw, circle, inner sep=1pt, fill=black] (P) at (1.1,0.5) {};
\node[above=2pt of P] {JANUS region};
\end{tikzpicture}
\caption{The stablecoin trilemma: It is challenging to optimize decentralization, capital efficiency, and safety/stability at the same time. JANUS strives to occupy a region closer to the triangle’s center, balancing all three dimensions.}
\label{fig:trilemma_visual}
\end{figure}
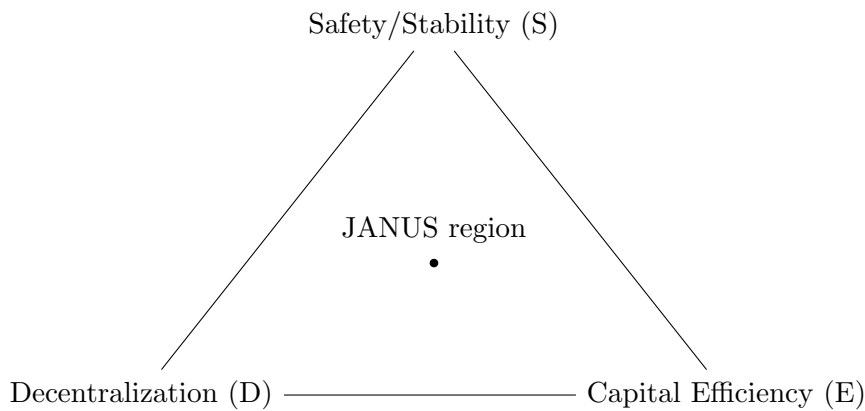

\paragraph{Defining Decentralization, Capital Efficiency, and Safety/Stability.}
In the Appendix, we present rigorous mathematical definitions for each dimension:
\begin{itemize}[noitemsep]
    \item $D(U)$: A measure of decentralization (e.g., via Herfindahl-like metrics of governance distribution). Higher $D(U)$ indicates more balanced control and minimal trust requirements.
    \item $E(U)$: A capital efficiency index, typically the ratio of stablecoin supply (at target price) to total collateral value. Lower overcollateralization implies higher $E(U)$.
    \item $S(U)$: Safety/stability modeled as $1 - \mathbb{P}(\mathcal{F})$, where $\mathbb{P}(\mathcal{F})$ is the probability of catastrophic failure, such as a prolonged peg break.
\end{itemize}

While no stablecoin design can perfectly maximize D, E, and S, JANUS’s dual-token model, multi-collateralization, soft peg, and AI-driven stabilization collectively \emph{expand the feasible frontier}, yielding higher $(D(U), E(U), S(U))$ than prior solutions. In the following sections, we examine how these components reduce systemic fragility and foster robust, inflation-adjusted value, culminating in a non-ponzinomic equilibrium.

\section{Core Architecture, Vault Ecosystem, and Stabilization Mechanisms}

Designing a stablecoin that addresses decentralization, capital efficiency, and safety/stability requires multiple synergistic components. JANUS adopts a dual-token approach, diversifies collateral sources, allows a soft peg to tolerate minor price deviations, and employs an AI-driven feedback loop to maintain equilibrium. This section outlines these key architectural elements, followed by an overview of the vault structures and stabilization strategies that support them.

\subsection{Core Design Elements: Dual-Token, Multi-Collateralization, and Soft Peg}
\paragraph{Dual-Token Architecture.}
JANUS introduces two primary tokens:
\begin{description}[noitemsep]
    \item[Alpha ($A$):] Influenced by crypto-market conditions, protocol governance, and fee mechanisms.
    \item[Omega ($\Omega$):] Partially backed by real-world asset (RWA) yields, providing a baseline value and reducing reliance on speculative inflows.
\end{description}
Omega’s external yield helps anchor the system and avoids the purely reflexive demand spiral seen in some single-token designs. Even in low-demand or bear market conditions, the protocol can sustain non-zero equilibria underpinned by external asset flows.

\paragraph{Multi-Collateralization with Crypto and RWAs.}
Depending solely on crypto collateral subjects a stablecoin to correlated market downturns, often necessitating heavy overcollateralization. JANUS integrates tokenized RWAs (e.g., treasury bills, trade finance receivables) alongside crypto assets, lowering systemic volatility and enhancing $S(U)$. By reducing correlation among collateral, the protocol also mitigates liquidation cascades, potentially minimizing the collateral ratio required for a given stablecoin supply and hence improving $E(U)$.

\paragraph{Soft Peg and Controlled Oscillations.}
JANUS defines a reference price $P_{\text{ref}}(t)$ that grows over time to account for inflation or desired appreciation. The system tolerates small deviations within:
\[
P_{\text{ref}}(1-\epsilon) \;\;\le\;\; P_{A,\Omega}(t) \;\;\le\;\; P_{\text{ref}}(1+\epsilon).
\]
Such flexibility prevents panic runs that often occur when a strict peg breaks slightly. This mechanism closely resembles managed exchange rates in certain macroeconomic policies, gently guiding prices back to equilibrium while avoiding severe interventions.

\subsection{Vault Ecosystem and AI-Driven Stabilization}
\paragraph{Vault Structures and Financial Instruments.}
Beyond simply depositing collateral to mint stablecoins, JANUS offers specialized vaults:
\begin{itemize}[noitemsep]
    \item \textbf{Genesis Vaults:} Users deposit collateral to mint Alpha/Omega at inception.
    \item \textbf{Borrow/Lending Vaults:} Participants lend tokens or borrow stable assets, increasing capital efficiency by deploying idle collateral. This fosters liquidity and reduces the need to liquidate holdings.
    \item \textbf{Fixed-Rate vs. Variable-Rate Vaults:} Some vaults provide predictable yields, while others dynamically adjust rates based on supply-demand fluctuations, aligning with more fluid market realities.
    \item \textbf{Infinite vs. Fixed Horizon Vaults:} Time-structured vaults (akin to bonds) coexist with perpetual structures, catering to diverse risk/return preferences.
\end{itemize}
Such diversity grants the AI controller multiple levers to maintain equilibrium, balancing user demand across a variety of instruments.

\begin{figure}[h!]
\centering
\begin{tikzpicture}[node distance=1.8cm, auto,>=latex', every node/.style={font=\small}]
\node[draw, rectangle, align=center] (crypto) {Crypto Collateral};
\node[draw, rectangle, align=center, right=1.5cm of crypto] (rwa) {RWA Collateral};
\node[draw, rectangle, align=center, below=1.5cm of $(crypto)!0.5!(rwa)$] (vault) {Genesis Vault \\(Mints Alpha/Omega)};
\draw[->] (crypto) -- (vault);
\draw[->] (rwa) -- (vault);
\node[draw, rectangle, align=center, below=1.5cm of vault] (borrow) {Borrow/Lending Vaults};
\node[draw, rectangle, align=center, below=1.5cm of borrow] (fixed) {Fixed-Rate / Variable-Rate Vaults};
\node[draw, rectangle, align=center, below=1.5cm of fixed] (horizon) {Infinite \& Fixed Horizon Vaults};
\draw[->] (vault) -- (borrow);
\draw[->] (borrow) -- (fixed);
\draw[->] (fixed) -- (horizon);
\end{tikzpicture}
\caption{Vault ecosystem: From genesis vaults that mint Alpha/Omega, participants access borrowing/lending layers, rate-structured vaults, and horizon-specific products, creating a robust financial environment for stability and adaptation.}
\end{figure}
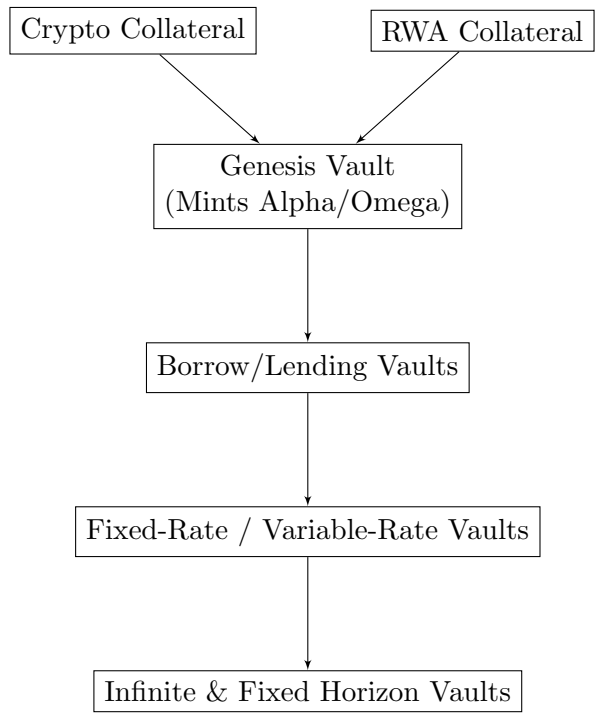

\paragraph{AI-Driven Feedback Loops.}
Central to JANUS’s stabilization lies an AI-driven mechanism that continuously monitors:
\begin{itemize}[noitemsep]
    \item $P_A(t)$ and $P_\Omega(t)$: token prices,
    \item collateral ratios and vault usage,
    \item liquidity conditions,
    \item volatility indicators and user demand.
\end{itemize}
When prices deviate from $P_{\text{ref}}(t)$, the AI modulates fees, rewards, or vault configurations. This negative feedback loop parallels certain central bank open market operations, but operates autonomously on the blockchain, ensuring minimal trust requirements.

\begin{figure}[h!]
\centering
\begin{tikzpicture}[node distance=2.3cm, auto,>=latex', every node/.style={font=\small}]
\node[draw, rectangle, align=center] (market) {Market Conditions \\($P_A,P_\Omega,\text{volatility}$)};
\node[draw, rectangle, align=center, right=3.5cm of market] (ai) {AI Controller};
\node[draw, rectangle, align=center, right=3.5cm of ai] (actions) {Actions: \\Adjust Rewards, Fees, Params};
\path[->] (market) edge (ai);
\path[->] (ai) edge (actions);
\draw[->] (actions.south) -- ++(0,-1.2) -| (market.south) node[midway,below]{System Response};
\end{tikzpicture}
\caption{AI-driven feedback loop: The AI observes market states, triggers parameter adjustments, and the new system response re-enters the loop, preserving stable, non-ponzinomic equilibria.}
\end{figure}
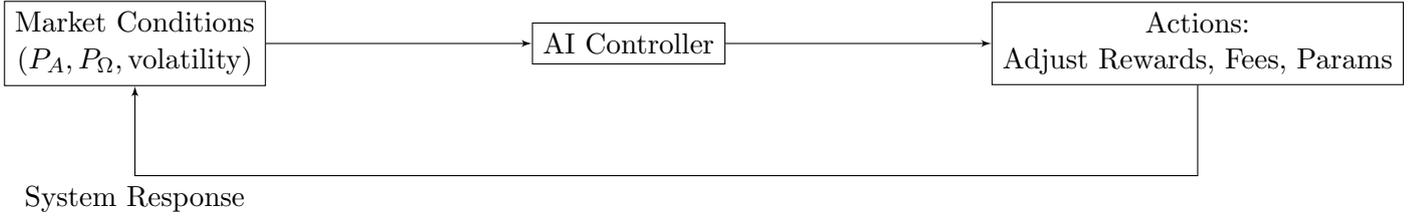

\begin{figure}[h!]
\centering
\includegraphics[width=0.8\textwidth]{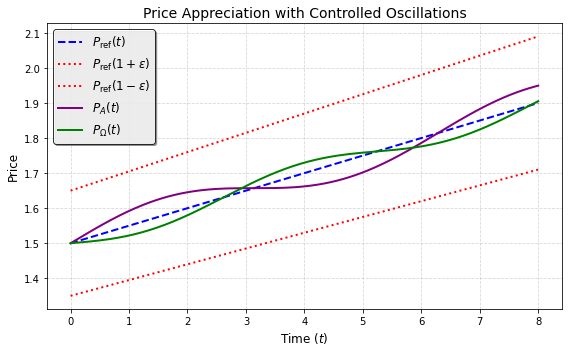}
\caption{Price appreciation with controlled oscillations. Both $P_A(t)$ and $P_\Omega(t)$ steadily appreciate over time (mirroring inflation-adjusted growth in $P_{\text{ref}}(t)$) while tolerating minor oscillations that enhance long-term stability.}
\label{fig:oscillations}
\end{figure}

\paragraph{Price Appreciation with Controlled Oscillations.}
In addition to the flowcharts above, Figure \ref{fig:oscillations} shows how Alpha and Omega can steadily appreciate over time while tolerating small oscillations around $P_{\text{ref}}(t)$. This approach preserves long-term inflation-adjusted value without risking severe peg breaks.

By integrating all these components—dual tokens, multi-collateralization, a soft peg with minor acceptable deviations, an AI-driven control loop, and diverse vault products—JANUS dynamically stabilizes under varying market scenarios. The system curtails catastrophic collapses (improving $S(U)$), avoids unnecessary overcollateralization (improving $E(U)$), and disperses governance to minimize trust points (improving $D(U)$).

\section{Implementation Strategies and Future Research Directions}
While the JANUS blueprint provides a robust theoretical foundation, practical deployment involves several key steps:
\begin{itemize}[noitemsep]
    \item \textbf{RWA Oracles and Legal Frameworks:} Ensuring transparent, legally compliant tokenized real-world assets to anchor Omega.
    \item \textbf{Distributed Governance:} Spreading governance tokens to enhance $D(U)$ and reduce centralized control or single points of failure.
    \item \textbf{Parameter Optimization:} Calibrating AI parameters, vault APYs, and fee structures for diverse market conditions, potentially using agent-based simulations.
    \item \textbf{Stress Testing:} Running scenario analyses (e.g., extreme crypto downturns, RWA yield shortfalls) to validate that negative feedback loops continue to maintain stability.
\end{itemize}

Future research may delve into:
\begin{itemize}[noitemsep]
    \item Machine learning approaches for predictive AI control, adjusting parameters proactively before major shocks.
    \item Game-theoretic analyses of participant incentives, especially underwriters, liquidity providers, and governance token holders.
    \item Expanding to an $N$-token ecosystem with specialized roles, from trade finance tokens to sustainability-linked assets, further diversifying risk and societal impact.
\end{itemize}

\section{Conclusion}
JANUS represents a new frontier in stablecoin design, weaving together dual tokens, multi-collateralization, a soft peg, and AI-driven stabilization to better satisfy the stablecoin trilemma. By introducing external yield sources (e.g., RWAs), distributing governance, and allowing mild price oscillations around an inflation-adjusted reference, JANUS simultaneously enhances decentralization, capital efficiency, and safety/stability.

Although no single solution can perfectly solve all stablecoin trade-offs, JANUS’s architecture takes a major step forward—offering a more resilient, inflation-aware, and globally relevant stablecoin. This synergy of economic theory, macroeconomic analogies, and cryptoeconomic design aims to provide robust, decentralized digital money adaptable to a rapidly evolving DeFi landscape.

\appendix
\section*{Appendix: Formal Definitions and Mathematical Treatment of the Trilemma}
This appendix details the mathematical underpinnings referenced throughout the main text, including definitions of decentralization, capital efficiency, and safety/stability, as well as equilibrium proofs and the transition from single-token to multi-asset systems.

\subsection*{Defining D(U), E(U), and S(U)}

\paragraph{Decentralization ($D(U)$).}
If $\omega_i$ is the governance fraction held by participant $i$ with $\sum_i \omega_i = 1$, one can define:
\[
D(U) = 1 - \sum_i \omega_i^2.
\]
A higher $D(U)$ indicates more distributed control (often measured via Herfindahl-based metrics). Additional trust-minimization factors (e.g., fully on-chain governance, multi-sig committees) can refine $D(U)$ further, but this baseline suffices for clarity.

\paragraph{Capital Efficiency ($E(U)$).}
Let $S_{\text{sc}}(t)$ be the stablecoin supply and $C_{\text{total}}(t)$ the total collateral value. We define:
\[
E(U) = \frac{S_{\text{sc}}(t) \,\cdot\, P_{\text{ref}}(t)}{C_{\text{total}}(t)}.
\]
High $E(U)$ means supporting a given stablecoin supply with less locked collateral, thereby boosting resource efficiency. Overly high collateral ratios harm $E(U)$ by underutilizing capital.

\paragraph{Safety/Stability ($S(U)$).}
Let $\mathbb{P}(\mathcal{F})$ be the probability of a catastrophic failure (e.g., a run or a prolonged peg break). We define:
\[
S(U)=1-\mathbb{P}(\mathcal{F}).
\]
Keeping $\mathbb{P}(\mathcal{F})$ near zero yields $S(U)$ close to 1, reflecting high resiliency under stress scenarios. Mechanisms such as external yield, multi-collateralization, and robust feedback loops help reduce $\mathbb{P}(\mathcal{F})$.

\subsection*{Existence and Stability of Equilibria}
Under continuous supply--demand and collateral constraints, we can build a mapping
\[
\mathbf{F}:\mathbf{x}\mapsto \mathbf{x}',
\]
where $\mathbf{x}$ is the system’s state vector (prices, collateral usage, vault distributions, etc.). By \emph{Brouwer’s fixed-point theorem}, a fixed point $\mathbf{x}^*$ exists, guaranteeing at least one equilibrium. Non-zero real-world yields ($r_{\text{RWA}}>0$) prevent a trivial zero-price solution, ensuring the protocol can anchor on fundamental value. Linearizing around $\mathbf{x}^*$ and examining the Jacobian reveals local stability if negative feedback loops dominate. In JANUS, these loops arise from AI-driven adjustments to fees, rewards, or vault configurations whenever token prices deviate from the reference band, thus lowering $\mathbb{P}(\mathcal{F})$ and raising $S(U)$.

\subsection*{Ponzinomic Pitfalls and Non-Ponzi Fundamentals}
A system is called \emph{ponzinomic} if token price appreciation depends solely on continuous new inflows, absent any external yield or uncorrelated asset backing. Formally, if the expected price $P_{\mathrm{token}}(t)$ satisfies:
\[
P_{\mathrm{token}}(t+1) \;=\; P_{\mathrm{token}}(t)\;+\;\delta_{\mathrm{inflow}}(t)\;-\;\delta_{\mathrm{outflow}}(t),
\]
where $\delta_{\mathrm{inflow}}(t)$ must remain positive and growing for $P_{\mathrm{token}}(t)$ not to collapse, the design is essentially a pyramid scheme. Once new buyers vanish, price collapses to zero, because there is no external source of value (like $r_{\text{RWA}}>0$ or uncorrelated asset reserves) to sustain it.

In contrast, JANUS ensures \emph{non-ponzinomic fundamentals} by introducing external yield (e.g., RWA) and uncorrelated asset backing. If one set of assets $\mathcal{A}_1$ (say, crypto collateral) has value $V_1$, and another $\mathcal{A}_2$ (say, RWA or any uncorrelated pool) has value $V_2$, the total support for minted tokens $T_i$ is at least $V_1 + V_2$—assuming no or low correlation among $\mathcal{A}_1$ and $\mathcal{A}_2$. Mathematically, if minted tokens have a notional value $M$, then for non-ponzinomic equilibrium, we generally require:
\[
M \;\le\; V_1 \;+\; \mathbb{E}[\,V_2 \,],
\]
where $\mathbb{E}[\,V_2\,]$ is the expected fundamental yield or value from the uncorrelated asset pool. As long as $M$ does not exceed the sum of these distinct value anchors, the system does not rely exclusively on new entrants to maintain token prices.

This principle extends to $N$ asset classes as well. Each new uncorrelated or low-correlation asset backing reduces overall variance and ensures that the protocol’s price appreciation does not hinge on a single continuous inflow. Instead, it can tap distinct yield sources or uncorrelated markets to provide a baseline. This not only increases $S(U)$ by lowering $\mathbb{P}(\mathcal{F})$ (the probability of a collapse) but also can improve $E(U)$ by letting the protocol sustain a higher token supply without pushing into unsustainable, reflexive territory.

\subsection*{From Single-Token to Multi-Asset, N-Token Systems}
A single-token stablecoin without external yield is prone to collapse whenever inflows slow, as no fundamental value halts a negative spiral. Adding a second token (Omega) anchored by real-world yield or uncorrelated asset classes adds a second dimension to absorb shocks. More generally, an $N$-token architecture diversifies risk further. 

If each token $T_i$ is backed by a different uncorrelated asset $C_i$, the total collateral $\mathrm{Var}(C_{\text{total}})$ can be reduced through low correlation $\rho_{ij}$:
\[
\mathrm{Var}(C_{\text{total}}) = \sum_i \theta_i^2\mathrm{Var}(C_i) 
+ 2\sum_{i<j} \theta_i\theta_j \rho_{ij}\sqrt{\mathrm{Var}(C_i)\,\mathrm{Var}(C_j)},
\]
where $\theta_i$ are collateral weights. Decreasing correlation lowers systemic fragility and thus $\mathbb{P}(\mathcal{F})$. In that sense, multi-asset, $N$-token systems that rely on externally anchored, uncorrelated assets move closer to satisfying the stablecoin trilemma of high $D(U), E(U), S(U)$.

\subsection*{References}
\begin{enumerate}[label={[}\arabic*{]}]
\item J. D. Farmer et al., “Stability and Instability in Algorithmic Stablecoins: Economic and Systemic Considerations,” Working Paper, 2023.
\item M. Raiden, “A Survey of Algorithmic Stablecoins,” \emph{arXiv preprint}, 20XX.
\item MakerDAO, “The DAI Stablecoin System,” \url{https://makerdao.com}.
\item Tether, “Transparency and Reserves,” \url{https://tether.to/en/transparency}.
\item A. Evans, “What’s Wrong with Algorithmic Stablecoins?” \emph{arXiv:2106.10551}, 2021.
\item Reflexer Labs, “RAI: Reflex-Indexed Stable Asset,” \url{https://reflexer.finance/}.
\item Centrifuge, “Real-World Asset Integration,” \url{https://centrifuge.io/}.
\item R. B. Mendoza and D. Y. Javier, “Inflation-Indexed Currencies and Stablecoins,” \emph{J. Monetary Economics}, 2022.
\item G. Angeris and T. Chitra, “Improved Price Oracles: Constant Function Market Makers,” ACM AFT, 2020.
\item B. Evans and K. Reed, “Algebraic Properties of Automated Market Makers,” Proc. ACM Meas. Anal. Comput. Syst. 5.2, 2021.
\item S. Karpus et al., “Stable Equilibria in DeFi Lending Pools,” \emph{arXiv preprint}, 2022.
\item M. Lyons and N. Khoja, “Stablecoins: Classifications, Market, and Regulatory Developments,” BIS Paper, 2022.
\item Bank for International Settlements (BIS), “Stablecoins: Risks, Potential and Regulation,” BIS AER, 2021.
\item K. C. Border, \emph{Fixed Point Theorems with Applications to Economics and Game Theory}, CUP, 1985.
\item K. J. Arrow and G. Debreu, “Existence of an Equilibrium for a Competitive Economy,” Econometrica 22(3), 1954.
\item G. Debreu, “Theory of Value,” Yale University Press, 1959.
\item P. Aghion and P. Howitt, “Endogenous Growth Theory,” MIT Press, 1998.
\item M. Obstfeld and K. Rogoff, “Foundations of International Macroeconomics,” MIT Press, 1996.
\item D. Acemoglu and A. Ozdaglar, “Machine Learning and Equilibria in Dynamic Environments,” \emph{Annual Rev. of Economics}, 12 (2020).
\end{enumerate}

\end{document}